\documentclass[12pt]{article}
\usepackage{epsfig,amssymb,amsmath,psfrag}

\def\tf {  \tilde{f}  }
\def\tC {  \tilde{C}  }

\newcommand{\cN}{{\cal N}}
\newcommand{\cR}{{\cal R}}
\newcommand{\cO}{{\cal O}}

\def \be  {\begin{equation}}
\def \ee  {\end{equation}}
\def \ba  {\begin{eqnarray}}
\def \ea  {\end{eqnarray}}

\newcommand \tbr [1] {\langle{#1}\rangle}
\newcommand \mfbr [1] {\langle{#1}\rangle}
\newcommand \fbr [1] {({#1})}

\begin{document}
\thispagestyle{empty}

\begin{flushright}
HU-EP-10/49\\
LAPTH-031/10
\end{flushright}

\begingroup\centering
{\Large\bfseries\mathversion{bold} Simple loop integrals and amplitudes in $\cN=4$ SYM \par}%
\vspace{8mm}

\begingroup\scshape\large 
James M.~Drummond,
\endgroup
\vspace{5mm}

\begingroup
\textit{LAPTH, Universit\'e de Savoie, CNRS\\
B.P. 110, F-74941 Annecy-le-Vieux Cedex, France }\par
\texttt{drummond@lapp.in2p3.fr\phantom{\ldots}}
\endgroup

\vspace{1cm}

\begingroup\scshape\large 
Johannes M.~Henn,
\endgroup
\vspace{5mm}

\begingroup
\textit{Institut f\"ur Physik, Humboldt-Universit\"at zu Berlin, \\
Newtonstra{\ss}e 15, D-12489 Berlin, Germany}\par
\texttt{henn@physik.hu-berlin.de\phantom{\ldots}}
\endgroup

\vspace{2cm}

\textbf{Abstract}\vspace{5mm}\par
\begin{minipage}{14.7cm}
We use momentum twistors to evaluate planar loop integrals.
Infrared divergences are regulated by the recently proposed
AdS-inspired mass regulator. We show that two-loop 
amplitudes in $\cN=4$ super Yang-Mills can be expanded
in terms of basis integrals having twistor numerators. We argue
that these integrals are considerably simpler compared to
the ones conventionally used.
Our case in point is the two-loop six-point MHV amplitude.
We present analytical results for the remainder function
in a kinematical limit, and find agreement with a recent
Wilson loop computation.
We also provide two-loop evidence that the logarithm
of MHV amplitudes can be written in terms of simple
twistor space integrals.
\end{minipage}\par
\endgroup 
\newpage

\tableofcontents

\setcounter{tocdepth}{2}

\newpage

\section{Introduction}

Recent years have seen tremendous progress in calculations
of loop level amplitudes in planar $\cN=4$ super Yang-Mills (SYM).
State-of-the-art methods based on generalised unitarity 
\cite{Bern:1994zx,Bern:1994cg} 
and sophisticated methods for the evaluation of loop integrals \cite{smirnov2006feynman}
have made the computation of various amplitudes at high numbers of
loops and with many external legs possible, 
see e.g. \cite{Anastasiou:2003kj,Bern:2005iz,Bern:2006ew,Spradlin:2008uu}.
Current techniques allow, for example, the numerical evaluation of six-point
MHV or NMHV amplitudes at two loops \cite{Bern:2008ap,Radu-talk}.
Obtaining analytical results for $n=6$ or computing $n >6$ amplitudes at two-loops 
however seems very involved using these methods.
On the other hand, recent encouraging results \cite{Goncharov:2010jf} suggest that one should try to find
analytical answers for amplitudes in $\cN=4$ SYM, and despite the progress 
mentioned above it seems clear that for doing
so we need new tools and insights.\\

In \cite{Alday:2009zm}, an alternative regularisation for infrared
divergences in planar $\cN=4$ SYM was put forward 
(see \cite{Schabinger:2008ah,McGreevy:2008zy} for earlier references).
It is inspired by the AdS description of scattering amplitudes \cite{Alday:2007hr}
and consists in considering scattering amplitudes on the
Coulomb branch of $\cN=4$ SYM.
The vacuum expectation values of some of the scalars 
give rise to masses that can be used to regulate the loop integrals.
It was shown that this has a number of conceptual as well as
practical advantages over dimensional regularisation, and
that it facilitates the computation of amplitudes with many 
loops or external legs \cite{Henn:2010bk,Henn:2010ir}.\\

In this paper we propose a new method that combines the virtues
of the mass regulator with those of momentum twistor variables \cite{Hodges:2009hk}\footnote{It is encouraging that the same variables were 
introduced independently at strong coupling \cite{Alday:2009yn}.}.
The latter are well-suited to describe planar four-dimensional
scattering amplitudes.
Recent papers discussed the one-loop box integrals using momentum
twistors \cite{Hodges:2010kq,Mason:2010pg}.
Here we apply momentum twistors to higher loop integrals.
We argue that integrals with certain momentum twistor
numerators are much simpler compared to the integrals that
are conventionally used\footnote{We are particularly grateful to Nima Arkani-Hamed, Jake Bourjaily, Freddy Cachazo, Simon Caron-Huot and Jaroslav Trnka for sharing their ideas related to integrals constructed using momentum twistors.}.
We show how to use twistor 
identities to express amplitudes quite generally in 
terms of our preferred basis.\\

We focus on amplitudes with $n \ge 6$ external particles.
The reason is that dual conformal symmetry, which is widely believed
to apply to planar scattering amplitudes in $\cN=4$ SYM, entirely fixes the
form of the four- and five-particle amplitudes, to any loop order \cite{Drummond:2007cf,Drummond:2007au}.
(This is related to a conjectured relationship between Wilson
loops and MHV scattering amplitudes \cite{Alday:2007hr,Drummond:2007aua, Brandhuber:2007yx, Drummond:2007cf}, for reviews see \cite{Alday:2008yw,Henn:2009bd}).
For $n\ge 6$ external legs there is the freedom of an arbitrary function, called
remainder function, that can depend on conformal cross-ratios only. Indeed the two-loop MHV amplitude and corresponding  Wilson loop have a non-trivial remainder function \cite{Drummond:2007bm,Bern:2008ap,Drummond:2008aq}.
As a specific example we then rewrite the six-point MHV 
amplitude in terms of our preferred basis integrals.
We evaluate the latter and compute the remainder 
function analytically in special kinematical regimes.\\

We are confident that the methods developed here and in our forthcoming paper \cite{diffeqpaper2}
can be used to make contact with recent results for certain two-loop Wilson loops
in various kinematical regimes, see \cite{Alday:2010ku} and  \cite{DelDuca:2010zp,Heslop:2010kq}. 
For example, it is straightforward to apply the methods presented here to
arbitrary $n$-point two-loop MHV amplitudes \cite{Vergu:2009tu}.
Moreover, while the correspondence with Wilson loops is, at the moment,
restricted to MHV amplitudes, the methods we develop in this paper 
are quite general and can also be used to study non-MHV amplitudes.\\

The outline of our paper is as follows: In section \ref{sect-intro-twistors}, we review the symmetries
of scattering amplitudes focussing on momentum
twistor variables and the massive regularisation. We give a one-loop example to
show how twistor numerators can easily be dealt with. We then go on to describe in
section \ref{sect-improved-basis} how to use twistor identities in order to transpose
an amplitude expressed in terms of canonical integrals in terms of our preferred basis.
We use the six-point two-loop MHV amplitude as an example, and discuss the virtues
of the new basis. In section \ref{sect-evaluation} we evaluate the integrals for the six-point two-loop MHV amplitude
and provide an analytical result for the remainder function in particular kinematical regimes.
In section \ref{sect-log-4pt} we provide another example for the usefulness of twistor variables
by showing that the logarithm of the two-loop four-point MHV amplitude can be written
as a simple twistor integral. In section \ref{sect-soft} we present a simple but useful consistency
test of loop integrands based in the soft limit.
Finally, in the appendix we present more technical results quoted in the main text, as well
as a proof that all leading singularities in $\cN=4$ SYM are dual conformal covariant.

\section{Lightning review of momentum twistor space and massive loop integrals}
\label{sect-intro-twistors}

Dual conformal symmetry is an important property of planar colour-ordered amplitudes in $\mathcal{N}=4$ super Yang-Mills theory. Let us introduce the basic notions.
Given the $n$ incoming light-like momenta of a planar ordered amplitude,
\be
p_i^{\alpha \dot \alpha} = \lambda_i^\alpha \tilde{\lambda}_i^{\dot \alpha}\,,
\ee
we define the dual coordinates in the usual manner \cite{Broadhurst:1993ib,Drummond:2006rz}, 
\be
x_i^{\alpha \dot\alpha} - x_{i+1}^{\alpha \dot\alpha} = p_i^{\alpha \dot\alpha}\,.
\ee
The dual $x_i$ define a light-like polygon in the dual space. Under conformal transformations of the dual coordinates (dual conformal transformations) the polygon transforms but the edges remain light-like. This fact means that one can discuss dual conformal transformations of planar ordered scattering amplitudes. In the $\mathcal{N}=4$ theory they are actually symmetries of the amplitudes, exact at tree-level \cite{Drummond:2008vq,Brandhuber:2008pf,Drummond:2008cr} but anomalous when acting on a suitably defined finite part at loop level \cite{Drummond:2007cf,Drummond:2007au,Drummond:2008vq}.\\

When taken together with the ordinary superconformal symmetry of the $\mathcal{N}=4$ super Yang-Mills Lagrangian, the dual conformal transformations generate the Yangian of the superconformal algebra, $Y(psl(4|4))$ \cite{Drummond:2009fd}. The Yangian is a symmetry of the tree-level amplitudes (with the usual caveats about contact terms in the action of the ordinary superconformal symmetry \cite{Bargheer:2009qu,Korchemsky:2009hm,Sever:2009aa}). At loop level the amplitudes break the ordinary superconformal symmetry, however all leading singularities (see \cite{Cachazo:2008vp,Cachazo:2008hp,ArkaniHamed:2009dn} for discussions of leading singularities) are ordinary superconformal invariants. This can be most easily seen in twistor space as they can be obtained by gluing tree-level amplitudes together in way which manifesty respects the superconformal symmetry \cite{Bullimore:2009cb,Kaplan:2009mh}. All leading singularities so far examined are also dual conformal. At one loop this essentially follows from the analysis in \cite{Drummond:2008bq,Brandhuber:2008pf}. We give a general proof of this fact for all leading singularities in appendix \ref{LSDCS}. This is relevant to the conjecture of \cite{ArkaniHamed:2009dn} that all leading singularities can be obtained from an integral over a Grassmannian of a particular kind. This dovetails very nicely with the fact that, under mild assumptions, the Grassmannian integral can be shown to give the most general form of a Yangian invariant, i.e. that all Yangian invaraints are obtained from the integral with some choice of contour \cite{Drummond:2010qh,Drummond:2010uq,Korchemsky:2010ut}.\\

Now let us turn to dual conformal symmetry realised at the level of the loop integrals. In \cite{Drummond:2006rz} it was noticed that the integrals appearing up to three-loops in the planar four-point amplitude formally exhibit a conformal symmetry in the dual space. Here we would like to discuss new integrals which also have this dual conformal property. Momentum twistors will be central to our discussion. These variables were introduced in \cite{Hodges:2009hk}. A very helpful discussion of the associated geometry is also given in \cite{Mason:2009qx}. They are the natural twistors associated with the dual coordinate space which can be used to describe scattering amplitudes.\\

A point in dual coordinate space corresponds to a (complex, projective) line in momentum twistor space. Two dual points are light-like separated if the corresponding lines in momentum twistor space intersect at some point in momentum twistor space. Thus the light-like polygon in dual space corresponds to a polygon in momentum twistor space with each line intersecting its two neighbouring lines as each dual point is light-like separated from its two neighbours.
The $n$ momentum twistors associated to this configuration of $n$ light-like lines are defined via the incidence relations,
\be
Z_i^A = (\lambda_i^\alpha , \mu_i^{\dot\alpha}), \qquad \mu_i^{\dot\alpha} = x_i^{\alpha \dot\alpha} \lambda_{i \alpha} = x_{i+1}^{\alpha \dot\alpha} \lambda_{i \alpha}\,.
\ee
The momentum twistor transforms linearly under the action of dual conformal symmetry, as indicated by the fundamental $sl(4)$ index $A$. Moreover the $n$ momentum twistors describing the polygon are free variables, in contrast to the dual points $x_i$ which obey the constraints of light-like separation from their neighbours. The dual point $x_i$ is associated the line described by the pair $Z_{i-1}^{[A} Z_i^{B]}$ or $(i-1 \,\, i)^{AB}$ for short. \\

Note that only the lines $(i-1 \,\,i)$ correspond to the cusps of the dual polygon. One can consider other lines in momentum twistor space, for example the line $(i \,\, i+2)$. This line obviously intersects the four lines $(i-1 \, i)$, $(i \, i+1)$, $(i+1 \, i+2)$ and $(i+2 \, i+3)$. It therefore corresponds to a point  which is light-like separated from the four points $x_i$, $x_{i+1}$, $x_{i+2}$ and $x_{i+3}$. This distinguishes it as one of the two solutions to the four-particle cut conditions when a one-mass box integral is thought of in dual coordinate language. The other solution we will denote as $\overline{(i \, i+1)}$ which is related to the first by parity. Both of the corresponding dual points are complex.\\

The incidence relations allow one to express functions of the $x_i$ in terms of momentum twistors. For example we have
\be
x_{ij}^2 = \frac{\fbr{i-1\,\, i\,\, j-1\,\, j}}{\langle i-1 \, i \rangle \langle j-1 \, j\rangle}\,,
\ee
where the four-brackets and two-brackets are defined as follows,
\be
\fbr{ijkl}  = \epsilon_{ABCD} Z_i^A Z_j^B Z_k^C Z_l^D, \qquad \langle i j \rangle = \lambda_i^\alpha \lambda_{j \alpha}\,.
\ee
The four-brackets are obviously dual conformal invariants while the two-brackets are invariant under just the Lorentz and (dual) translation transformations.\\

Since momentum twistors linearise the action of dual conformal symmetry it is clear that they are natural variables for discussing dual conformal integrals.
Let us discuss the formulation of loop integrals using momentum twistors. Recent papers have already employed these variables to discuss the one-loop box integrals \cite{Hodges:2010kq,Mason:2010pg}. Here we will describe how other integrals can be similarly formulated using momentum twistors. In particular we are interested in integrals which make explicit use of the twistor lines which do not correspond to the dual $x_i$ like the line $(i\,\,i+2)$ described above. As an example we will consider a one-loop pentagon integral,
\be
I_5 = \int \frac{d^4 Z_{AB}}{i \pi^2} \frac{ \fbr{AB13}\fbr{4512}\fbr{2345}}{\fbr{AB12}\fbr{AB23}\fbr{AB34}\fbr{AB45}\fbr{AB51}}\,.
\label{twistorpentagon}
\ee

 \begin{figure}[t]
 \psfrag{1}[cc][cc]{$Z_{1}$}
\psfrag{2}[cc][cc]{$Z_{2}$}
\psfrag{3}[cc][cc]{$Z_{3}$}
\psfrag{4}[cc][cc]{$Z_{4}$}
\psfrag{5}[cc][cc]{$Z_{5}$}
\psfrag{x1}[cc][cc]{$x_{1}$}
\psfrag{x2}[cc][cc]{$x_{2}$}
\psfrag{x3}[cc][cc]{$x_{3}$}
\psfrag{x4}[cc][cc]{$x_{4}$}
\psfrag{x5}[cc][cc]{$x_{5}$}
\psfrag{xa}[cc][cc]{$x_{a}$}
 \centerline{
 {\epsfxsize12cm  \epsfbox{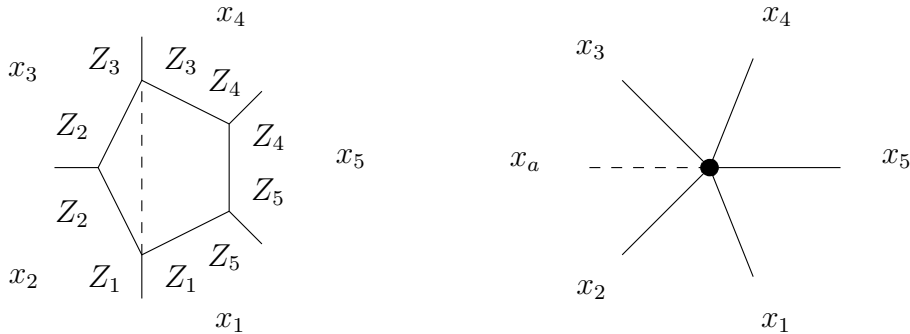}} 
}
\caption{\small
Pentagon integral (\ref{twistorpentagon}) with twistor numerator $\fbr{13AB}$. The figure on the right
shows the same integral in the more familiar dual space notation, with the dashed line denoting a 
numerator connecting to the complex point $x_{a}$, see (\ref{dualxpentagon}).
}
\label{fig-canonicalbasis}
\end{figure}

Here the integration is over the space of lines in momentum twistor space $(AB)$. As we have discussed, this is equivalent to an integration over points in dual space. Recall that the momentum twistors are only defined up to scaling so it is important that the above expression has zero scaling weight in all of the momentum twistor variables. Taking into account that the integration measure
\be
d^4Z_{AB} \sim d^4 x_0 \langle AB \rangle^4\,,
\ee
has scaling weight 4 for the twistors $A$ and $B$, we see that (\ref{twistorpentagon}) indeed has zero scaling weight. The numerator of the integrand in (\ref{twistorpentagon}) contains the factor $(AB13)$ depending on the line $(13)$ which, as we have discussed corresponds to a complex point.\\

The integral form in (\ref{twistorpentagon}) is manifestly dual conformal invariant, being written in terms of four-brackets, even though we are describing a pentagon integral. In dual coordinate notation the pentagon integral is
\be
I_5 = \frac{x_{52}^2 x_{35}^2}{x_{5a}^2} \int \frac{d^4 x_0}{i \pi^2} \frac{x_{0a}^2}{x_{01}^2 x_{02}^2 x_{03}^2 x_{04}^2 x_{05}^2}\,,
\label{dualxpentagon}
\ee
where the point $x_a$ is one of the two solutions to
\be
x_{5a}^2 = x_{1a}^2= x_{2a}^2 =x_{3a}^2 = 0\,.
\label{defxa}
\ee
We can see from (\ref{dualxpentagon}) that the integral form is dual conformal invariant, having dual conformal weight zero at all of the points $x_1,x_2,x_3,x_4,x_5,x_a$. The existence of a sixth point $x_a$ was crucial in being able to make a dual conformal pentagon integral. It is not independent of the other five dual points $x_1,\ldots,x_5$, being defined as a solution to the (one-mass box) cut-conditions (\ref{defxa}), nor is it real. The integral (\ref{twistorpentagon}) is therefore not invariant under parity, i.e. it is chiral.\\

The integral (\ref{twistorpentagon}) naturally leads us to introduce another tool that we will need in this paper. It is infrared divergent so we will need a regulator. We would like to preserve the dual conformal symmetry so we will use the AdS regularisation introduced in \cite{Alday:2009zm}. For actual calculations we will use the regularisation where all masses are equal. In particular this means the outermost propagators in the planar loop integrals that we are studying are modified as follows,
\be
\frac{1}{x_{ij}^2} \longrightarrow \frac{1}{x_{ij}^2 + m^2}\,.
\ee

In our example of the dual conformal one-loop pentagon this means that the five propagators need to be altered to incorporate the mass terms. In momentum twistor language this can be achieved by turning each of the pairs of twistors associated with the dual $x_i$ into a bitwistor with a component proportional to the infinity twistor $I^{AB}$. Thus we make the replacements in the denominator of (\ref{twistorpentagon})
\be
Z_{i-1}^{[ A} \, Z_{i}^{B ]} \longrightarrow X^{AB}_{i-1,i} = Z_{i-1}^{[A} \, Z_{i}^{B]} + m^2 \tbr{ i-1 \,\, i} I^{AB} \,,
\ee
where $[\cdot,\cdot]$ stands for the antisymmetric part.
This has the effect that each of the propagator factors becomes
\be
\frac{1}{ \fbr{ABi-1\,\,i}} \longrightarrow \frac{1}{\mfbr{ AB i-1 \,\, i}} = \frac{1}{\fbr{AB X_{i-1 \,i}}} =\frac{1}{\fbr{AB i-1 \,\,i} + m^2 \tbr{ AB } \tbr{ i-1 \,\, i}}\,.
\ee
We will leave the numerator of (\ref{twistorpentagon}) unchanged. In fact the definition of the numerator factor does not matter in this case as any potential modification would be $O(m^2)$ and the integral is only logarithmically divergent so these terms can be safely ignored for the present purpose.
Thus we arrive at the final definition for the regularised pentagon integral,
\be 
I_5 = \int \frac{ d^4Z_{AB}}{i\pi^2} \frac{ \fbr{AB13}\fbr{4512}\fbr{2345}}{\mfbr{ AB 12}\mfbr{ AB 23 } \mfbr{ AB 34 } \mfbr{ AB 45 } \mfbr{ AB 51 }}\,.
\ee
Let us proceed to find a Feynman parametrisation of this integral. In order to do this it is helpful to normalise the denominator factors so that they carry no scaling weights in the external twistor variables. We will define
\be
\hat{X}^{AB}_{i,i+1} = \frac{X^{AB}_{i,i+1}}{\tbr{i\,i+1}} = \frac{Z_{i}^{[A} \, Z_{i+1}^{B]} }{\tbr{i\,i+1}} + m^2 I^{AB} \,.
\ee
Then the integral can be written
\be
I_5 = \frac{\fbr{4512} \fbr{2345}}{\prod_i \tbr{i\, i+1}} \int \frac{d^4 Z_{AB}}{i \pi^2} \frac{ \fbr{AB13} }
{\fbr{AB\hat{X}_{12}} \fbr{AB\hat{X}_{23}}\fbr{AB\hat{X}_{34}}\fbr{AB \hat{X}_{45}} \fbr{AB\hat{X}_{51}}}\,.
\ee
Introducing Feynman parameters we can write this as
\be
I_5 = \frac{4! \fbr{4512}\fbr{2345}}{\prod_i \tbr{i \, i+1}} \int \frac{d^4 Z_{AB}}{i \pi^2} \int D^4 \alpha \frac{\fbr{AB13}}{\fbr{AB Y}^5}\,,
\ee
where the bitwistor $Y$ is defined by
\be
Y = \alpha_{12} \hat{X}_{12} + \alpha_{23} \hat{X}_{23} + \alpha_{34} \hat{X}_{34} + \alpha_{45} \hat{X}_{45} + \alpha_{51} \hat{X}_{51}\,.
\ee
We write the measure of the Feynman parameter integration as $D^4\alpha$ to remind ourselves that this is a projective integral.
The numerator factor can now neatly be written as a derivative,
\be
I_5 = - \frac{3! \fbr{4512}\fbr{2345}}{\prod_i \tbr{i\,i+1}} \int D^4 \alpha \fbr{13\partial_Y} \int  \frac{d^4Z_{AB}}{ i \pi^2} \frac{1}{\fbr{ABY}^4}\,.
\ee
The integral over the line $(AB)$ can now be performed as described in \cite{Hodges:2010kq,Mason:2010pg} with the result
\be
I_5 = -4 \frac{\fbr{4512}\fbr{2345}}{\prod_i \tbr{i\,i+1}}\int D^4 \alpha \fbr{13 \partial_Y} \frac{1}{\fbr{YY}^2} = 16\frac{\fbr{4512}\fbr{2345}}{\prod_i \tbr{i\,i+1}}\int D^4 \alpha \frac{\fbr{13Y}}{\fbr{YY}^3} \,.
\ee
The remaining four-brackets in the numerator and denominator can now be expressed in terms of the twistor variables,
\begin{align}
\fbr{13Y} &= \alpha_{45} \biggl[\frac{\fbr{1345}}{\tbr{45}} + m^2 \tbr{13}\biggr], \label{I5num}\\
\fbr{YY} &= 2\Bigl[ \sum_i \alpha_{i-1 \,i} \alpha_{i+1\,i+2} x_{i,i+2}^2  + \sum_i \alpha_{i\,i+1} m^2 \Bigr] \,.
\end{align}
Finally combining everything and ignoring the $O(m^2)$ term from the numerator in (\ref{I5num}) we obtain
\be
I_5 = -2 x_{41}^2 x_{52}^2 x_{35}^2 \int \frac{d^5 \alpha \,\delta\bigl( \sum_i \alpha_{i\,i+1} - 1\bigr) \alpha_{45}}
{\bigl[ \sum_i \alpha_{i-1 \,i}\alpha_{i+1\,i+2} x_{i,i+2}^2 + m^2 \bigr]^3} +  \cO(m^2)\,.
\ee
So in summary we see that the twistor numerators can be easily dealt with.
In the next section we are going to explain how such integrals can be
used in order to simplify expressions for loop integrands.

\section{New representations of loop integrands}
\label{sect-improved-basis}
In the previous sections we reviewed how momentum twistors and a massive infrared 
regulator can be used to discuss planar loop integrals. We also showed that 
certain non-local twistor numerators can easily be treated in an approach using
Feynman parameters. We will now explain how to use these non-standard
numerators in order to derive simpler expressions for the integrand of multi-loop
and multi-leg amplitudes.\\

As a case in point we are going to discuss the two-loop six-point MHV amplitude.
The form of the two-loop six-point MHV amplitude given by Bern et al \cite{Bern:2008ap} 
and transposed to the massive regularisation is
\begin{equation}\label{MHV2old}
M_{6}^{(2)} = \frac{1}{16}\, \sum_{ 12 \; {\rm perms/flips}} \, \sum_{i=1}^{13} \,c_{i}\, I_{6;2;i}+ \cO(m^2) \,,
\end{equation}
with the integrals $I_{6;2;i}$ correspond to the $I^{(i)}$ of \cite{Bern:2008ap}, except 
that we include a normalisation factor in order to make them dimensionless,
and the coefficients $c_{i}$, given in \cite{Bern:2008ap}, are modified accordingly.
\footnote{In \cite{Bern:2008ap} there are also two further integrals 
depending explicitly on $\mu$, the $(-2\epsilon)$-dimensional component(s) of the
loop momenta in dimensional regularisation.
It has been observed that these terms cancel to $\cO(\epsilon)$ in the $\log M_{6}$
to the two-loop level, which suggests that analogous terms, if present in the mass
regularisation, should be $\cO(m^2)$ \cite{Henn:2010ir}.
}
The integrals appearing in (\ref{MHV2old}) are of the double box, pentabox and double 
pentagon type, representatives of which are shown in Figure \ref{fig-canonicalbasis}. 
The latter two involve certain numerator factors that depend on
the loop momentum. Since they are more complicated than the double box
integrals, we wish to eliminate them in favour of simpler integrals.
We will now explain how this can be done quite generally.\\

 \begin{figure}[t]
 \psfrag{1}[cc][cc]{$I_{6;2;1}$}
\psfrag{5}[cc][cc]{$I_{6;2;5}$}
\psfrag{9}[cc][cc]{$I_{6;2;9}$}
\psfrag{12}[cc][cc]{$I_{6;2;12}$}
 \centerline{
 {\epsfxsize14cm  \epsfbox{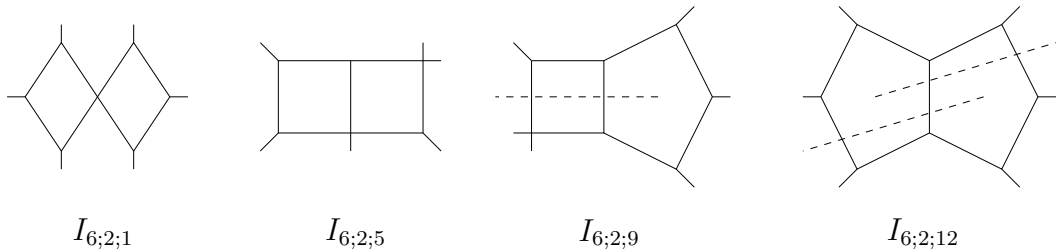}} 
}
\caption{\small
Representative integrals appearing in two-loop MHV amplitudes. In the pictures
the specific choice $n=6$ was made. The dashed lines stand for numerator factors
that depend on the loop momentum.
}
\label{fig-canonicalbasis}
\end{figure}

\begin{figure}[t]
 \psfrag{sum}[cc][cc]{$=$}
 \centerline{
 {\epsfxsize14cm  \epsfbox{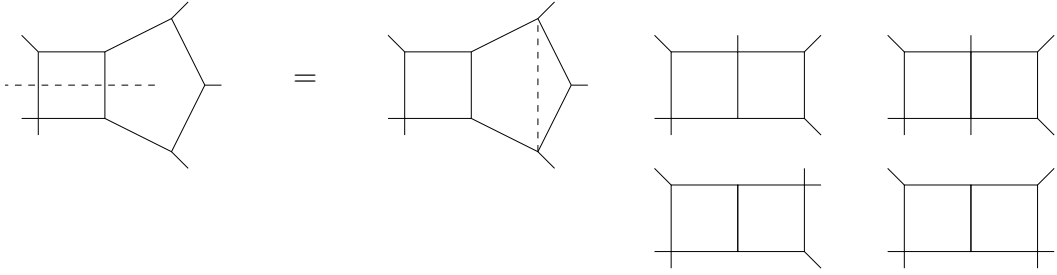}} 
}
\caption{\small
Example of identity (\ref{expansion}) when used to expand $I_{6;2;9}$ in terms
of our preferred basis. We do note display the (loop-momentum independent) prefactors and normalisations of the integrals.
}
\label{fig-loop2exidentity}
\end{figure}

Consider for example the following six-point pentabox integral,
\be
I_{6;2;9} \propto \int \frac{dZ^{AB} dZ^{CD} \mfbr{AB12}}{ \mfbr{AB23}\mfbr{AB34}\mfbr{AB45}\mfbr{AB56}\fbr{ABCD}\mfbr{CD56}\mfbr{CD12}\mfbr{CD23} } \,,
\ee
which is shown in Fig. \ref{fig-canonicalbasis}.
We would like to exchange the numerator $\mfbr{AB12}$ for terms which either cancel a propagator or give a preferred `non-local' type numerator of the form $\fbr{AB35}$ or $\fbr{AB\overline{35}}$. We can do this by writing
\be
(12) = b_{23} (23) + b_{34} (34) + b_{45} (45) + b_{56} (56) + b_{35} (35) + b_{\overline{35}} (\overline{35})\,.
\label{expansion}
\ee
Here
\be
(\overline{35})^{AB} = (234\,\cdot\,)^A(\,\cdot\,  456)^B - (A,B) \,.
\ee
By projecting with different twistors we can solve for the coefficients. We obtain
\begin{align}
&b_{23} = \frac{\fbr{1245}}{\fbr{2345}}, \quad b_{34} = \frac{\fbr{6245}\fbr{5312}}{\fbr{6345}\fbr{2345}}, \quad b_{45} = -\frac{\fbr{6235}\fbr{1234}}{\fbr{6345}\fbr{2345}} \,, \notag \\
&b_{56} = -\frac{\fbr{1234}}{\fbr{6345}}, \quad b_{35} = \frac{\fbr{6245}\fbr{1234}}{\fbr{4635}\fbr{2345}}, \quad b_{\overline{35}}=\frac{\fbr{1235}}{\fbr{6345}\fbr{2345}} \,.
\label{bcoefs}
\end{align}
Thus we have expressed the pentabox integral $I$ in terms of double box integrals and pentabox integrals with the preferred numerators $\fbr{AB35}$ and $\fbr{AB\overline{35}}$, up to $+\cO(m^2)$ terms that
arise due to the difference between $\mfbr{\ldots}$ and $\fbr{\ldots}$.
Moreover it turns out that the integral with $\fbr{AB\overline{35}}$ in the numerator is equivalent, again
to $\cO(m^2)$, to the $\fbr{AB35}$ integral. The resulting identity is schematically shown in 
Figure \ref{fig-loop2exidentity}.\\

One can use (\ref{expansion}) and (\ref{bcoefs}) and their reflected versions and similar identities obtained upon rotation to expand any unpleasant numerator.
For example if we consider the double pentagon integral with numerator $\mfbr{AB12}\mfbr{CD34}$ 
(with $AB$ and $CD$ representing the loop integration variables) we can expand it by using (\ref{expansion}) for the first factor and the similar identity obtained by reflection and rotation for the second factor.
The above are examples of a completely general identity which can be used for any such integral. 
It can be found in the Appendix, and can be used to simplify any of the pentaboxes or double pentagons appearing in any two-loop amplitude.\\

There is another type of term in the two-loop amplitudes that we would like
to remove, namely the ``kissing box'' topology, which are a product of two
one-loop box integrals, see $I_{6;2;1}$ in figure \ref{fig-canonicalbasis}. 
Of course these are analytically quite simple, but
we observe that they come with certain prefactors such that they 
contain rational factors, a feature which we would like to eliminate.
We can rewrite those integrals as a double-pentagon integral
by multiplying their integrand with $1 = \fbr{ABCD} /\fbr{ABCD}$,
and use the following identity to decompose the $\fbr{ABCD}$ in
the numerator:
\begin{eqnarray} \label{twistor-identity}
 \fbr{ ikAB } \fbr{ jlCD} + \fbr{ jlAB }\fbr{ikCD }  &=& 
\fbr{ ijAB }\fbr{klCD } +\fbr{ klAB }\fbr{ ijCD }  \nonumber \\
&& \hspace{-3cm} -  \fbr{ jkAB }  \fbr{ liCD } - \fbr{ liAB }  \fbr{ jkCD }   -  \fbr{ ijkl} \fbr{ ABCD } \,.   \end{eqnarray}
The identity holds for any $i,j,k,l$. The integrals we obtain
in this way can in turn be reduced using (\ref{expansion}), as described above.\\

\begin{figure}[t]
 \psfrag{6m1}[cc][cc]{$I_{6;2;m1}$}
 \psfrag{6m2}[cc][cc]{$I_{6;2;m2}$}
 \psfrag{6m3}[cc][cc]{$I_{6;2;m3}$}
 \psfrag{7}[cc][cc]{$I_{6;2;7}$}
 \psfrag{15}[cc][cc]{$I_{6;2;15}$}
 \centerline{
 {\epsfxsize14cm  \epsfbox{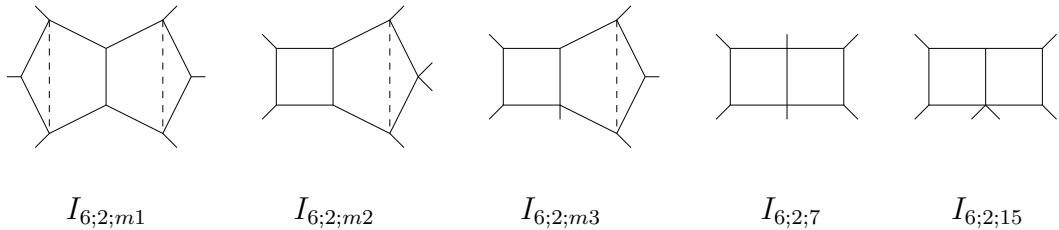}} 
}
\caption{\small
Integrals appearing in the improved representation of the six-point two-loop MHV amplitude, c.f. (\ref{MHV2loopimproved}). The dashed lines indicate non-local twistor numerators, and a normalisation
factor is not shown in the figure. Both are given explicitly in the Appendix.}
\label{fig-improved-basis}
\end{figure}

The numerator identities described above can be used to obtain
an improved representation of the two-loop six-point MHV amplitude,
which is given by \cite{refNima}
\begin{equation}\label{MHV2loopimproved}
M_{6}^{(2)} = \frac{1}{4} \sum_{ 12 \; {\rm perms/flips}} \left[ \frac{1}{4} I_{6;2;m1}  - \frac{1}{2} I_{6;2;m2} - \frac{1}{4} I_{6;2;m3} + \frac{1}{4} I_{6;2;7} + \frac{1}{2} I_{6;2;15}   \right] + \cO(m^2) \,,
\end{equation}
where all integrals are depicted in Figure \ref{fig-improved-basis}, 
and their explicit definitions including their normalisation is given in the Appendix.
The new representation (\ref{MHV2loopimproved}) is manifestly more compact
than (\ref{MHV2old}), as it requires only $5$ as opposed to $13$ integrals.
Moreover, it has a number of other virtues, as we discuss presently:
\begin{itemize}
\item
We find that all integrals appearing in (\ref{MHV2loopimproved}), to the order that we
evaluated them analytically, do not contain any rational functions multiplying the (poly-)logarithms
appearing in them. Experience shows that this is a generic property of scattering amplitudes in
$\cN=4$ SYM, and it is remarkable that it holds for all integrals in our basis, even before
summing them up.
This is not the case for the integrals appearing in (\ref{MHV2old}), see e.g. the expression
given for $I_{6;2;11}$ i.e. $I^{(11)}$ in Appendix A.2 of \cite{Bern:2008ap}.
\item
The previous point hints at a hidden simplicity of the integrals appearing in (\ref{MHV2loopimproved}). 
This can be explained in part by a new method based on differential equations \cite{Drummond:2006rz,diffeqpaper1} that will be presented in
a forthcoming paper \cite{diffeqpaper2}.
\item 
It has become standard to evaluate loop integrals using Mellin-Barnes representations \cite{smirnov2006feynman}.
We remark that the bottleneck of the calculation of $n\ge 6$ two-loop amplitudes used to be
the double pentagon integrals, as e.g. $I_{6;2;12}$ and $I_{6;2;13}$ in  (\ref{MHV2old}). 
This becomes clear from inspecting the dimensionality of the Mellin-Barnes representations
used to evaluate these integrals: the ones used for those two integrals in \cite{Bern:2008ap} are $18$-fold.
In the improved representation (\ref{MHV2loopimproved}), we need $12$-fold Mellin-Barnes
integrals at most.\footnote{The dimensionality of the Mellin-Barnes integrals is further reduced
considerably when taking the regulator limit, i.e. $\epsilon \to 0$ in dimensional regularisation, or
$m^2 \to 0$ here.}
\end{itemize}
Two remarks are in order:
Firstly, the method we described here to simplify the representations for loop
amplitudes/integrands applies more generally. For example, it can be used to simplify 
the expressions for $n$-point two-loop MHV integrand \cite{Vergu:2009tu}, the recently presented
form of the six-point two-loop NMHV integrand \cite{Radu-talk}, or any other loop amplitudes.
Secondly, we have already argued that the expressions we gave for the  six-point
two-loop MHV integrand greatly facilitates its evaluation, and we will see in the next section 
that this is indeed the case.
However it may very well be that ultimately there exist yet better representations
(from the point of view of evaluating the integrals), that can be obtained for example by
replacing the remaining double box integrals $I_{6;2;7}$ and $I_{6;2;15}$ by pentabox integrals
with ``magic'' numerators (other than the ones already appearing in $I_{6;2;m2}$  and $I_{6;2;m3}$).
We leave this question for future work.

\section{Two-loop amplitudes in $\cN = 4 $ SYM}
\label{sect-evaluation}
Here we discuss the evaluation of the loop integrals in the new basis described in the previous section.
As a specific example, we will present analytical results for the six-point two-loop amplitude
in certain kinematical limits. We stress that the methods presented here can be straightforwardly 
used to evaluate two-loop amplitudes with more external legs or different helicity configurations.

\subsection{Structure of the loop corrections for MHV amplitudes}

Let us begin by recalling the expression for the one-loop six-point amplitude.
Its integral representation is \cite{Bern:1994zx}
\begin{equation}\label{oneloopMHV6}
M_{6}^{(1)} = - \frac{1}{4} \sum_{6 \,{\rm perms}} \left[ F^{1m} - \frac{1}{2} F^{2me} \right] +\cO(m^2) \,.
\end{equation}
and using the explicit expressions for the integrals in equations (\ref{app-F1m}) and (\ref{app-F2me}) we see
that it is given by (dropping $\cO(m^2)$ terms)
\begin{eqnarray}\label{oneloopMHV6explicit}
M_{6}^{(1)} &=& - \frac{1}{4}  \sum_{i=1}^{6} \left[ \log^2 \frac{ m^2}{x_{i,i+2}^2} \right]  + F_{6}^{(1)} \,, \\
F_{6}^{(1)}  &=&  \frac{\pi^2}{2}  +\frac{1}{2}  \sum_{i=1}^{6} \left[ 
 - \log \frac{x_{i,i+2}^2}{x_{i,i+3}^2} \log \frac{x_{i+1,i+3}^2}{x_{i,i+3}^2}  +\frac{1}{4} \log^2 \frac{x_{i,i+3}^2}{x_{i+1,i+4}^2} -\frac{1}{2} {\rm Li}_{2}\left( 1- \frac{x_{i,i+2}^2 x_{i+3,i+5}^2}{x_{i,i+3}^2 x_{i+2,i+5}^2} \right) \right]  \,, \nonumber
\end{eqnarray}
where we used the identity $ {\rm Li}_{2}(1-x) +  {\rm Li}_{2}(1-1/x)+ 1/2 \log^{2} x = 0$. 
We remind the reader that the function $F_{6}^{(1)}$ satisfies the anomalous Ward identities of \cite{Drummond:2007au}.\\

We now wish to investigate this amplitude at the two-loop order. It is worthwhile 
to discuss the expected structure of $n$-point MHV amplitudes at higher loops.
In reference \cite{Henn:2010ir} it was suggested that 
the logarithm of the loop corrections to MHV amplitudes
in the massive regularisation should have the following form,
\begin{eqnarray}
\label{BDSnparticle}
\log M_n &=& \sum_{i=1}^n
\left[ - \frac{\gamma(a)}{16} 
\log^2 \left( \frac{x_{i,i+2}^2}{m^2} \right)
- \frac{\tilde{\cal G}_0(a)} {2}
\log\left( \frac{x_{i,i+2}^2}{m^2} \right)
+ \tf (a)
\right] \nonumber \\
&& + \frac{1}{4} \gamma(a) \, F_n^{(1)} + \cR_{n}(u_{i},a) + \tC (a)
+ \cO(m^2)\,.
\end{eqnarray}
This is motivated by the 
ABDK/BDS ansatz \cite{Anastasiou:2003kj,Bern:2005iz},
allowing for a function $\cR_{n}$ of conformal cross ratios.
Let us briefly discuss this formula.
Firstly, it states that the infrared divergent part of the amplitudes, which manifests itself
through logarithms in $m^2$ as $m^2 \to 0$, should exponentiate,
and that this exponentiation is governed by certain anomalous dimensions\footnote{The notation 
$\gamma(a) = 2 \, \Gamma_{\rm cusp}(a)$ is also widely used 
in the literature to denote the cusp anomalous dimension \cite{Korchemskaya:1992je}.} 
$\gamma(a)$ and $\tilde{\cal G}_0(a)$, in accord with \cite{Korchemsky:1988hd,Mitov:2006xs}.
The latter take the values
\begin{equation}
\gamma(a)  =  4 a - 4 \zeta_{2} a^2 + {\cal O}(a^3) \,,
\qquad\qquad 
\tilde{\cal G}_0(a) = - \zeta_{3} a^2 + {\cal O}(a^3) \,.
\end{equation}
Secondly, the finite terms in (\ref{BDSnparticle}) can be thought of as an explicit solution to the
Ward identities of \cite{Drummond:2007au}, with  $\frac{1}{4} \gamma(a) \, F_n^{(1)}$ being a
particular solution.
The full finite part allowed by the Ward identities is the particular solution plus a function of the available conformal cross-ratios $\cR_n(u_i)$.
It vanishes for $n=4$ and $n=5$, since non-vanishing 
conformal cross-ratios appear only starting at $n=6$.
Thirdly, the specific choice of
the kinematic-independent terms $\tf(a)$ and $\tC(a)$, determined by the four-
and five-point case, was made in such a way that the remainder function 
vanishes in the collinear limit.
They are given by \cite{Henn:2010ir}
\begin{equation}
\label{Cdef}
\tf(a)  =  \frac{\zeta_4}{2} a^2 + {\cal O}(a^3) \,,
\qquad\qquad 
\tC(a) = - \frac{5 \zeta_4}{4} a^2 + {\cal O}(a^3) \,.
\end{equation}
Specialising to the six-point case at two loops, we see that
$F_{6}^{(1)}$ was defined in the second line of (\ref{oneloopMHV6explicit}) and $\cR_{6} (a)=  \cO(a^2)$ by definition.
As we discussed, equation (\ref{BDSnparticle}) implies that all logarithms in $m^2$ of $M_{6}^{(2)}$ are captured by
\begin{equation}\label{M62divpart}
\left( \frac{1}{2} M_{6}^{(1)}  \right)^2 +  \sum_{i=1}^6
\left[ - \frac{\gamma^{(2)}}{16} 
\log^2 \left( \frac{x_{i,i+2}^2}{m^2} \right)
- \frac{\tilde{\cal G}_0^{(2)}} {2}
\log\left( \frac{x_{i,i+2}^2}{m^2} \right)
\right]\,.
\end{equation}
Moreover, we have that
\begin{equation}\label{definitionR6}
\cR_{6}^{(2)}(u_{1},u_{2},u_{3}) = M_{6}^{(2)} - {\rm Eqn.}\, (\ref{M62divpart}) - \frac{1}{4} \gamma^{(2)} \, F_6^{(1)} - \frac{7}{12} \zeta_{4}\,,
\end{equation}
where 
\begin{eqnarray}
u_{1} = \frac{ x_{13}^2 x_{46}^2} {x_{14}^2 x_{36}^2} \,,\quad 
u_{2} = \frac{ x_{24}^2 x_{15}^2} {x_{25}^2 x_{14}^2} \,,\quad 
u_{3} = \frac{ x_{25}^2 x_{26}^2} {x_{36}^2 x_{25}^2} \,,
\end{eqnarray}
are the three non-vanishing cross-ratios at six points.
In the following section, we will compute the loop integrals
contributing to $M_{6}^{(2)}$ and by virtue of (\ref{definitionR6}) obtain a
result for $\cR_{6}^{(2)}$.

\subsection{Evaluation of two-loop integrals and amplitudes}

Let us explain the general strategy for computing the integrals
contributing to the two-loop amplitude.
As was explained in section \ref{sect-improved-basis}, the twistor numerators can be conveniently 
dealt with, and the introduction of Feynman parameters for the propagators in the
usual way allows to carry out the loop integration.
It is then straightforward to write down Mellin-Barnes representations for all integrals.
The $m^2 \to 0$ limit can be implemented by {\tt MBasymptotics} \cite{heptools1}, and
{\tt barnesroutines} \cite{heptools2} can be used to further simplify the resulting expressions
by using Barnes lemmas where applicable. 
The resulting expressions are of similar complexity as those obtained
in the Wilson loop computation of \cite{DelDuca:2010zg} (recall that MHV amplitudes are
conjectured to be dual to $n$-cusped light-like Wilson loops).
One may therefore hope to derive analytical answers for the
remainder function, for example by exploiting that the cross-ratios
that the latter depends on are invariant under certain Regge limits,
as was successfully done in \cite{DelDuca:2010zg}.
A full analytical evaluation of the remainder would go beyond the
scope of this paper.
Here, we content ourselves to provide analytical
results for the six-point remainder function in certain kinematical limits\footnote{A new method for obtaining analytical results for loop integrals 
based on differential equations \cite{Drummond:2006rz,diffeqpaper1} will be presented in
a forthcoming paper \cite{diffeqpaper2}.}.
\\

The Mellin-Barnes representations we obtain can also be readily evaluated
numerically. It is amusing to note that the double-pentagon integrals of \cite{Bern:2008ap},
which constitute the bottle-neck of the calculations in the conventional
basis, have been replaced by simpler penta-box and double pentagon integrals,
and the finite double pentagon integral with ``magic'' numerator, which is 
much better behaved (In fact the latter integral is infrared finite. It would not
be an exaggeration to  
say that the momentum twistors allow us to replace the most difficult integral in the 
conventional representation by the simplest integral in the new representation).
Therefore we are in a position to produce high-accuracy numerical results
for the remainder function at arbitrary (Euclidean) values of the kinematical variables,
using e.g. the {\tt MB} package \cite{Czakon:2005rk}.
We remark that it is straightforward to extend this statement to other two-loop
amplitudes, e.g. of higher multiplicity or for other helicity configurations, since
the contributing loop integrals are expected to be of the same type.
Therefore in particular we have also numerical access to non-MHV amplitudes\footnote{
Preliminary results for the six-point NMHV amplitude in dimensional 
regularisation where presented in \cite{Radu-talk}.}, which
was not possible previously as the duality with Wilson loops is restricted
to MHV amplitudes so far.\\

Let us now return to our specific six-point MHV example. 
In section \ref{sect-improved-basis} we have already alluded to the fact that 
apart from containing fewer contributing integrals, the new representation of the loop amplitudes
also has other virtues. 
In order to see this,
let us evaluate the integrals in the improved representation of section \ref{sect-improved-basis}.
To begin with, we verified the infrared structure (\ref{M62divpart}).
The reader is invited to check this by the explicit formula we give for the $\log^{4} m^2 $
and $\log^3 m^2 $ terms of the integrals in the appendix. 
We also computed analytically the  $\log^2 m^2$ terms but
refrain from writing out the lengthier formulas.\\

For simplicity, let us now consider the symmetric kinematical configuration
where $x_{i,i+2}^2 = 1$ and $x_{i,i+3}^2 = u^{-1/2}$, which sets all cross-ratios
equal $u_{i}=u$. 
As an illustration of the above statements about the infrared structure of 
the amplitude, let us quote the result for the kinematical point $u=1$,
for which we obtain (denoting $L = \log m^2 $)\footnote{All coefficients were obtained analytically, except the constant in $I_{6;2;m1}$, which was obtained to high numerical accuracy.}
\begin{eqnarray}
I_{6;2;m1} &=& -6 \zeta_{4} + \cO(m^2) \,,\\
I_{6;2;m2} &=& -\frac{1}{24} L^4 - \zeta_{2} L^2 -8 \zeta_{3} L -11\zeta_{4} + \cO(m^2) \,,\\
I_{6;2;m3} &=&  \cO(m^2) \,,\\
I_{6;2;15} &=& \frac{7}{12} L^4 - 4 \zeta_{2} L^2 -16 \zeta_{3} L -12 \zeta_{4} + \cO(m^2) \,,\\
I_{6;2;7} &=& \frac{1}{4} L^4 +2 \zeta_{2} L^2 +12 \zeta_{3} L +12 \zeta_{4} +\cO(m^2)\,.
\end{eqnarray}
At the symmetric point (\ref{MHV2loopimproved}) becomes
\begin{eqnarray}
M_{6}^{(2)} &=& \frac{1}{4}  \left[ 3  I_{6;2;m1}  - 6 I_{6;2;m2}   - 3 I_{6;2;m3} + 3  I_{6;2;7} + 6  I_{6;2;15} \right]   \,, \\
&=&  \frac{9}{8} L^4 - \frac{7}{8} L^2 - 3 \zeta_{3} L + \frac{27}{160} \pi^4   + \cO(m^2) \,.
\end{eqnarray}
and we find perfect agreement with the expected infrared structure according to equation (\ref{M62divpart}).
Moreover, from (\ref{definitionR6}) we see that
\begin{equation}
\cR_{6}^{(2)}(1,1,1) =- \frac{\pi^4}{36}\,,
\end{equation}
in perfect agreement with the corresponding analytical result obtained in \cite{DelDuca:2010zg} for the dual Wilson loop.
We also evaluate the limits where $u \to 0$ and $u \to \infty$ analytically, and obtain (the intermediate results
are given in the Appendix)
\begin{equation}
\lim_{u\to 0} \, \cR_{6}^{(2)}(u,u,u) = \frac{\pi^2}{8} \log^2 u +\frac{17 \pi^4}{1440} + \cO(u) \,,
\end{equation}
and
\begin{equation}
\lim_{u\to \infty} \, \cR_{6}^{(2)}(u,u,u) = -\frac{\pi^4}{144} + \cO(1/u) \,,
\end{equation}
again in perfect agreement with the Wilson loop result of \cite{DelDuca:2010zg}.\\

In summary, the new representation of the loop integrand, combined with the virtues of the regulator of \cite{Alday:2009zm},
has allowed us to easily produce an analytical result for the six-point remainder function (in a kinematical limit).
It seems straightforward to evaluate other two-loop amplitudes in a similar way.

\section{Twistor space integral for exponentiated four-point two-loop amplitude}
\label{sect-log-4pt}

In the previous sections we argued that momentum twistors can
be used to simplify the integral representations for two-loop amplitudes.
In particular, we exploited a new form of the two-loop six-point MHV
amplitude to evaluate the remainder function and obtained analytical
results in certain kinematical limits.\\

Here we wish to focus on MHV amplitudes and give a further application
of momentum twistors.
From the general structure of infrared divergences, see equation (\ref{BDSnparticle}),
it is clear that a good quantity to study is the logarithm of the loop
corrections $M_{n}$ of MHV amplitudes. 
We recall that the mass regulator as such is already convenient for this purpose,
as it allows to switch between $M_{n}$ and $\log M_{n}$ without having to keep
$\cO(m^2)$ terms\footnote{In dimensional regularisation, $\cO(\epsilon)$ terms are required
for this.}.
However, the $L$-loop amplitude $M_{n}^{(L)}$ starts with $\log^{2L} m^2 $ as most divergent 
term, whereas in $\log M_{n}$ the most divergent term is $\log^2 m^2$, and so
many terms must cancel between different loop orders when taking
the logarithm.
One can ask whether an integral representation exists that directly yields this logarithm,
without having to compute those higher terms in the first place?\\

As we show presently for the four-particle amplitude at two loops, this is indeed the case.
There exists a natural way to present $\log M_{4}$ that has the above properties.
As we will see, momentum twistors will again be crucial.\\

To begin with, we recall the expression for the four-point amplitude to two loops,
\begin{eqnarray}
 \left( \log M_{4} \right)^{(2)} &=& \frac{1}{4} \left[  I_{2}(s,t,m^2) + I_{2}(t,s,m^2) - \frac{1}{2} \left( I_{1}(s,t,m^2) \right)^2  \right] \label{two-loop-exp} 
\end{eqnarray}
where $I_{1}$ and $I_{2}$ are the one- and two-loop ladder integrals, respectively, and $s=x_{13}^2$ and $t=x_{24}^2$.
The result for the small $m^2$ expansion is in agreement with (\ref{BDSnparticle}) for $n=4$, and is explicitly 
given by \cite{Alday:2009zm}
\begin{eqnarray}
 \left( \log M_{4} \right)^{(2)}  &=& \zeta_{2}  \log \frac{m^2}{s}  \log \frac{m^2}{t}  - \zeta_{3} \left( \log \frac{m^2}{s} + \log \frac{m^2}{s} \right) - \frac{3}{40} \pi^4 + \cO(m^2) \,. \label{two-loop-exp2} 
\end{eqnarray}
The idea is now to combine all integrals on the r.h.s. of (\ref{two-loop-exp})  
and to simplify the resulting expression using momentum twistors.
At this point we recall an identity that was given in section \ref{sect-improved-basis},
\begin{eqnarray} \label{twistor-identity2}
 \fbr{ 13AB } \fbr{ 24CD} + \fbr{ 24AB }\fbr{13CD }  &=& 
\fbr{ 12AB }\fbr{34CD } +\fbr{ 34AB }\fbr{ 12CD }  \nonumber \\
&& \hspace{-3cm} -  \fbr{ 23AB }  \fbr{ 41CD } - \fbr{ 41AB }  \fbr{ 23CD }   -  \fbr{ 1234 } \fbr{ ABCD } \,.   \end{eqnarray}
We see that, after symmetrising the integrand in the integration variables
the coefficients of the terms exactly such that we can use (\ref{twistor-identity2})!
In doing so, we have to take into account that all of the propagators in the integrals except the middle
line of the double ladders contain a $+m^2$ term. Let us introduce the notation $(1234) = \langle 1234 \rangle + m^2 \langle12 \rangle 
\langle34\rangle$ and
\begin{eqnarray}
I \lbrack Q \rbrack := \int \frac{dZ_{AB} dZ_{CD}}{(i \pi^2)^2} \frac{Q \, \fbr{ 1234}^3 }{ \prod_{i=1}^{4}\left[  \mfbr{ {i-1,i} AB )( {i-1,i} CD } \right]  \fbr{ AB CD } }\,,
\end{eqnarray}
where the cyclicity of the indices is understood and $Q$ can be a loop-dependent numerator that is to be included in the
integration.
Then, we can write
\begin{eqnarray} \label{magiclogm4}
 \left( \log M_{4} \right)^{(2)} =  \frac{1}{8} \, I \lbrack \mfbr{ 13AB} \mfbr{24CD} +  m^2 \fbr{ 1234} \tbr{ AB }\tbr{ CD }  \; + AB \leftrightarrow CD  \rbrack \,.
\end{eqnarray}
This is the main result of this section. The logarithm of the four-particle amplitude at two loops can essentially
be written as a single twistor integral, with ``magic'' numerator $(13AB)  (24CD)$. 
In what follows we will verify explicitly that (\ref{magiclogm4})
gives the correct answer (\ref{two-loop-exp2}). As advertised, we will not have to compute 
any terms higher than $\log^2 m^2$.\\

A word of caution is due with respect to the $m^2  \fbr{1234} \tbr{AB} \tbr{CD}$ term
in the numerator -- usually the integrals we deal with have only logarithmic divergences in $m^2$ as $m^2 \to 0$,
so that such terms remain $\cO(m^2)$ after integration and can be neglected. Here however, writing all terms over 
a common denominator has increased the number of propagators and lead to a more singular denominator. 
As a result this term has to be kept, as we will see.\\

We now compute the small $m^2$ expansion of the two integrals in (\ref{magiclogm4}). 
Proceeding as explained in section \ref{sect-intro-twistors} we find the following Feynman parameter formula for 
$I_{2}\lbrack  \mfbr{13AB}\mfbr{24CD} \rbrack$,
\begin{eqnarray}
 2 s^2 t^2 \int_{0}^{\infty} \frac{ d^{4}\beta_{i} \delta( \sum_{i=1}^{4} \beta_{i} -1)}{ (\beta_{1} \beta_{3} s + \beta_{2} \beta_{4} t + m^2 )}  \int_{0}^{\infty} \frac{  d^{5}\alpha_{i} \delta( \sum_{i=1}^{5} \alpha_{i} -1) \alpha_{5} }{ (\gamma_{1} \gamma_{3} s + \gamma_{2} \gamma_{4} t + m^2 )^{3} } + \cO(m^2) \,,
\end{eqnarray}
where $\gamma_{i} = \alpha_{5} \beta_{i} + \alpha_{i}$.
The small $m^2$ limit is then readily evaluated using Mellin-Barnes methods, with the (remarkably simple) result
\begin{eqnarray}\label{result1}
4 \zeta_{2} \log \frac{m^2}{s} \log \frac{m^2}{t} + (-4 \zeta_{3} + 8 \zeta_{2} ) \left( \log \frac{m^2}{s}  + \log \frac{m^2}{t}  \right) - 
17.768
+ \cO(m^2)\,.
\end{eqnarray}
The $m^2 \fbr{1234} \tbr{AB} \tbr{CD}$ term can be evaluated in a similar way. We find
\begin{equation}\label{subtle-integral}
  -8 \zeta_{2}\left( \log \frac{m^2}{s}  + \log \frac{m^2}{t}  \right)  -11.454 + \cO(m^2) \,.
\end{equation}
Combining equations (\ref{result1}) and (\ref{subtle-integral}) according to (\ref{magiclogm4}) we have
 complete agreement with equation (\ref{two-loop-exp2}).
It is remarkable that no $\log^4 m^2$ terms appeared in the calculation, and no cancellations of say ${\rm Li}_{4}$ terms were required.\\

In conclusion, we found a remarkably simple integral representation of $\log M_{4}$ at the two-loop level.
The fact that we were able to use the twistor identity (\ref{twistor-identity}) relied on the relative coefficient
between the one-loop and two-loop integrals, which in turn is fixed by the infrared structure of the amplitude.
This may suggest that similar formulas can be written at higher loops or for higher multiplicities.

\section{Consistency of the loop integrand with soft limits}
\label{sect-soft}

It is well known that collinear and soft limits can provide an important
consistency check for loop amplitudes \cite{Bern:1994zx}.
Here we propose that the soft limit can also be used
in a more direct way as a consistency check of the loop {\it integrand}.
This check works in a very simple, almost pictorial way, and does
not require any loop integrals to be evaluated.\\

How do the scattering amplitudes behave in the soft limit, where 
where the momentum of one of the scattered particles is sent to zero, e.g. $p_{n} \to 0$?
The generic structure is, e.g. at the one-loop level \cite{Bern:1994zx},
\begin{equation}\label{soft1}
\lim_{p_{n} \to 0}  M_{n} \to M_{n-1} + div \,, 
\end{equation}
where the $div$ is some universal soft divergence 
We argue that the divergent term in (\ref{soft1}) arises
due to the non-commutativity of two limits:
the soft limit and the regulator limit (e.g. 
$\epsilon \to 0$ in dimensional or $m^2 \to 0$ in
mass regularisation).
For finite value of the regulator, there is no
reason for a divergent term to appear, and we expect
the soft limit to be completely smooth, 
\begin{equation}\label{soft2}
\lim_{p_{n} \to 0} M_{n} \to M_{n-1} \,,
\end{equation}
to all loop orders.\\

How will we be able to use (\ref{soft2}) in practice? 
We would like to argue that (\ref{soft2}) can be a useful
guiding principle for constructing the correct loop integrand.
Indeed if we can find an ansatz for the amplitude that manifestly respects
(\ref{soft2}) at the level of the integrand, then it will trivially also be
true for the integrated amplitude.\\

Let us now give some examples to see the usefulness of this approach.
Our first example is the one-loop MHV amplitude $M_{n}$. 
Let us consider the ansatz
\begin{equation}\label{oneloop-npointMHV}
M_{n} =  \frac{1}{4} \, \sum_{a,b} F_{2me}(x_{a},x_{a+1},x_{b},x_{b+1}) \,,
\end{equation}
where the sum runs over all inequivalent unordered pairs $(a,b)$.
This is indeed the correct answer \cite{Bern:1994zx}, at least up to $+ \cO(m^2)$ terms, which
may in particular contain parity odd contributions.
Note that some of the ``two-mass easy'' integrals in (\ref{oneloop-npointMHV}) 
are in fact ``one-mass'' integrals (for specific values of the labels),
but we shall leave this distinction implicit.
Let us now analyse the consistency of  (\ref{oneloop-npointMHV})  with the soft limit 
(\ref{soft2}).
The integrand of a generic two-mass easy function $F_{2me}$ is given by
\begin{eqnarray}\label{F2me-integrand}
{
 \psfrag{xa}[cc][cc]{$x_{a+1}\;$}
 \psfrag{xap}[cc][cc]{$x_{b}$}
 \psfrag{xb}[cc][cc]{$\; x_{b+1}$}
 \psfrag{xbp}[cc][cc]{$x_{a}$}
\psfrag{dots}[cc][cc]{$\dots$}
\parbox[c]{30mm}{\includegraphics[height = 25mm]{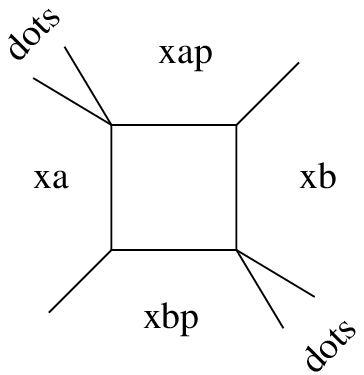}} 
}
\longleftrightarrow  \;  \;  \frac{x_{ab}^2 x_{a+1,b+1}^2 -x_{b+1,a}^2 x_{a+1,b}^2 }{(x_{a,i}^2 + m^2)(x_{a+1,i}^2 + m^2)(x_{b,i}^2 + m^2)(x_{b+1,i}^2 + m^2)}  \,,
\end{eqnarray}
where $x_{i}$ is the integration variable. 
We see that there are two different possibilities: 
in the limit where a momentum belonging to a massive corner
vanishes, we simply obtain lower-point integral $F_{2me}$, 
with the unchanged labels.
On the other hand, if the momentum taken to zero is $p_{a}$ or
$p_{b}$, the numerator in (\ref{F2me-integrand}) makes the expression vanish. This is important,
because otherwise we would have obtained an unwanted triangle integral with doubled propagator.
So we see that one-loop MHV amplitudes (\ref{oneloop-npointMHV}) manifestly satisfy (\ref{soft2}).\\

Our second example is the two-loop MHV amplitude,
which works in exactly the same way.
Since the improved representation uses twistor variables,
we have to formulate the soft limit in terms of those variables.
When momentum $p_{n}$ goes to zero, we have that
the twistor $Z_{n}$ becomes a linear combination of $Z_{1}$ and $Z_{n-1}$,
i.e. 
\begin{equation}
Z_{n} \to \alpha Z_{1} + \beta Z_{n-1} \,.
\end{equation}
The interested reader is invited to verify that the integrand of
the five-particle amplitude can be recovered in this way.\\

We remark that the way the MHV amplitudes manage to be manifestly consistent
with (\ref{soft2}) is very reminiscent of how the same soft limit works for tree-level amplitudes,
as was discussed in \cite{Drummond:2008cr}. So far we only discussed the soft limit that
does not change the helicity class of the amplitude. It is expected that the soft limit
that goes from an $n$-point N${}^{k}$MHV amplitude to an $(n-1)$-point N${}^{k-1}$MHV amplitude
is also smooth. This should provide a useful consistency check on the loop integrand of non-MHV 
amplitudes.\\

We hope to have convinced the reader of the usefulness of this test. A word of caution
is due: it is well-known that the soft/collinear limit cannot completely
constrain the amplitude, as there can be non-trivial functions that vanish in all
limits, the remainder function $\cR_{6}$ being a good example. Similarly,
one can easily think of integrands that vanish in all soft limits due to numerator factors,
such as e.g. $\fbr{1234} \fbr{3456} \fbr{5612}$ in the six-point case.

\section*{Acknowledgements}
We are grateful to N.~Arkani-Hamed for several invitations to the Institute for Advanced
Study, where part of this work was done. We are indebted to N.~Arkani-Hamed,
J.~Bourjaily, F.~Cachazo, S.~Caron-Huot and J.~Trnka for numerous enlightening discussions,
and for discussing the contents of \cite{refNima} with us before publication.
J.~H. thanks Z.~Bern for interesting discussions and encouragement.
\appendix

\section{Dual conformal symmetry of leading singularities}
\label{LSDCS}
Here we give a proof that all leading singularities in planar $\mathcal{N}=4$ super Yang-Mills theory are dual conformal covariant. The calculation proceeds as follows. First we will start with an arbitrary planar graph constructed from tree-level amplitudes glued over the internal edges. If there are $4l$ cut propagators then the fact that the internal lines are on shell cut propagators means that the loop integration is localised by these conditions to a discrete set of possibilities of which one must be chosen. If it there are fewer than $4l$ cut propagators then one must specify further the nature of the contour - there will be other poles around which the loop integration must be taken - if not then the leading singularity would be zero. Such leading singularities are called {\sl composite} leading singularities. Once the contour has been completely specified as a closed contour integral all internal momenta are fixed in terms of the external ones. 

The vertices of the graph correspond to tree-level superamplitudes of $\cN=4$ super Yang-Mills theory. These tree-level amplitudes can be of any helicity type (MHV, NMHV, etc) including the three-point $\overline{\rm MHV}$ amplitude. Following \cite{Bullimore:2009cb} we will replace any non-MHV amplitude with its BCF expression, e.g. for an NMHV tree-level amplitude we will write it as a sum over the gluings of four tree-level vertices of lower MHV degree. In this way we will increase the number of loops but will reduce every graph to a sum over graphs with only MHV and $\overline{\rm MHV}_3$ vertices. These graphs are called {\sl primitive} in the language of \cite{Bullimore:2009cb}.

Now that the graphs have been reduced to ones involving only MHV and $\overline{\rm MHV}_3$ vertices, the next step is to perform the sum over exchanged states on each internal leg. In other words we have to perform the Grassmann integral associated to each internal leg. The Grassmann integrations can be performed via three different operations which we will call A,B and C:

{\bf A. $\overline{\rm MHV}_3$ glued to MHV:}

The relevant Grassmann integral is
\be
\int d^4 \eta_l \delta^4(\eta_l [12] + \eta_1[2l] + \eta_2 [l1]) \delta^8(-\lambda_l \eta_l + q_K) = [12]^4 \delta^8(q).
\ee
In other words as far as the Grassmann factors are concerned one can replace the $\overline{\rm MHV}_3$- MHV${}_{n-1}$ pair with a single MHV${}_n$ vertex.

{\bf B. MHV glued to MHV:}

Here the relevant Grassmann integral is
\be
\int d^4\eta_l \delta^8(q_{K_1} + \lambda_l \eta_l) \delta^8(-\lambda_l \eta_l + q_{K_2}) = \delta^8(q) \delta^4(\langle l q_{K_1} \rangle) = \delta^8(q) \delta^4(\langle l q_{K_2} \rangle) 
\ee
Thus one obtains an MHV vertex with a decoration factor of the form $\delta^4(\langle l q_{K_1} \rangle)$.
Under the delta function $\delta^8(q)$ we can introduce a dual notation by writing 
\be
q_{K_1} = \psi_1 - \psi_2 = - q_{K_2}.
\ee
One of the $\psi$ variables will be identified with a external dual $\theta$ variable if one or other side of the gluing line corresponds  to the outside of the original graph. Otherwise they are just shorthand for a given linear combination of supercharges, e.g. $q_{K_1}$ which may contain Grassmann variables that still have to be integrated over. Writing the decoration factor in terms of the $\psi$ variables it takes the form
\be
\delta^4(\langle l \psi_1\rangle - \langle l \psi_2 \rangle).
\ee

The decoration factors can then go on take part in further Grassmann integrations. This brings us to the final operation,

{\bf C. MHV doubly glued to MHV:}
After several applications of operations A and B one will be left with a loop that connects two MHV vertices with some extra dependence on the integration variables in factors of the form $\delta^4(...)$.

Let us consider the example
\begin{align}
&\int d^4 \eta_{l_1} d^4 \eta_{l_2} \delta^8(q_{K_1} + q_{K_2} + \lambda_{l_1} \eta_{l_1} + \lambda_{l_2} \eta_{l_2}) \delta^4(\langle l \psi_l\rangle - \langle l \psi_{l_i} \rangle) \delta^8(-\lambda_{l_1} \eta_{l_1} - \lambda_{l_2} \eta_{l_2} + q_{K_3}) \notag \\
&= \delta^8(q) \int d^4 \eta_{l_1} d^4\eta_{l_2} \delta^4(\langle l \psi_l\rangle - \langle l \psi_{l_i} \rangle) \delta^8(-\lambda_{l_1} \eta_{l_1} - \lambda_{l_2} \eta_{l_2} + q_{K_3}).
\label{Cintegral}
\end{align}
Here the $\psi$ variables are defined so that $\psi_l - \psi_{l_i} = q_{K_2} + \lambda_{l_1} \eta_{l_1}=  -q_{K_1} - \lambda_{l_2} \eta_{l_2}$. If we further define $\psi_{l_1}$ and $\psi_{l_2}$ via 
\be
\psi_{l_1} - \psi_{l_2} = q_{K_3}, \qquad \psi_l - \psi_{l_1} = q_{K_2}
\ee
then we have that 
\be
\langle l_1 \psi_{l_i} \rangle = \langle l_1 \psi_{l_1} \rangle, \qquad \langle l_2 \psi_{l_i} \rangle = \langle l_2 \psi_{l_2}\rangle.
\ee
Hence we can reexpress $\psi_i$ in terms if the `external' $\psi$ variables,
\be
\psi_i  = \frac{\langle l_1 \psi_{l_1}\rangle \lambda_{l_2} - \langle l_2 \psi_{l_2} \rangle \lambda_{l_1}}{\langle l_1 l_2 \rangle}.
\label{psirep}
\ee
The final factor in (\ref{Cintegral}) then just contributes $\langle l_1 l_2 \rangle^4$ and we obtain 
\be
\delta^8(q) \delta^4\bigl( \langle l \psi_l \rangle \langle l_1 l_2 \rangle + \langle l_1 \psi_{l_1} \rangle \langle l_2 l \rangle + \langle l_2 \psi_{l_2} \rangle \langle l l_1 \rangle \bigr).
\label{3ptdeco}
\ee
Thus we see that the decoration factor develops more terms in the delta function.

This calculation easily generalises to more complicated decoration factors. For example let us imagine that we have already performed one loop of Grassmann integrations. Then we will obtain a decoration factor of the type in (\ref{3ptdeco}). When this appears under another integration of type C we obtain
\begin{align}
\int d^4\eta_{l_1} d^4 \eta_{l_2} &\delta^8(q_{K_1}+q_{K_2} + q_{K_3} + \lambda_{l_1} \eta_{l_1} + \lambda_{l_2} \eta_{l_2}) \notag \\
& \delta^4 \bigl(\langle l_a \psi_{l_a} \rangle \langle l_b l_i \rangle + \langle l_b \psi_{l_b} \rangle \langle l_i l_a \rangle + \langle l_i \psi_{l_i}  \rangle \langle l_a l_b \rangle \bigr) \delta^8(-\lambda_{l_1} \eta_{l_1} - \lambda_{l_2} \eta_{l_2} + q_{K_4}).
\end{align}
The integral can be performed just as before using (\ref{psirep}) as we obtain
\be
\delta^8(q) \delta^4\bigl(\langle l_a \psi_{l_a} \rangle \langle l_b l_i \rangle \langle l_1 l_2\rangle + \langle l_b \psi_{l_b} \rangle \langle l_i l_a \rangle \langle l_1 l_2 \rangle + \langle l_1 \psi_{l_1} \rangle \langle l_i l_2 \rangle \langle l_a l_b \rangle + \langle l_2 \psi_{l_2} \rangle \langle l_1 l_i \rangle \langle l_a l_b \rangle \bigr).
\ee

Generically there will be several decoration factors all depending on the internal $\psi$ variable. In this case all of them develop more terms under the Grassmann integration. 

Once all Grassmann integrations have been performed one can replace the external $\psi$ variables with dual $\theta$ variables. The resulting expression is then clearly dual conformal covariant because every dual $\theta$ is only ever contracted with the neighbouring $\lambda$ from the original diagram and $\lambda$ variables are only ever contracted with neighbouring $\lambda$ variables.

The remaining step is simply to count the dual conformal weights of every factor we have produced throughout the calculation. Doing do one finds that they combine with the weights of the bosonic factors in precisely the right way.
This also shows that all contributions have the same weight and so the result for the leading singularity is dual conformal covariant.

\section{Two-loop integrals}

\subsection{Twistor identity to relate numerator factors}
In section \ref{sect-improved-basis} we gave an example for an identity that
can be used to express pentabox and double pentagon integrals in terms
of our preferred basis. The identity shown there is a special case a more 
general identity,
\be
(l \, l+1) = a_{j1}(j-1\, j) + a_{j2} (j \, j+1) + a_{k1} (k-1 \, k) + a_{k2} (k \, k+1) + a_{jk} (jk) + a_{\overline{jk}} (\overline{jk}) \,,
\ee
where 
\be
(\overline{jk})^{AB} = (j-1\,j\,j+1 \,\, \cdot)^A (\cdot \,\, k-1 \,k \,k+1)^{B} - (B,A) \,.
\ee
The expansion coefficients are 
\be
a_{xy} = \frac{\hat{a}_{xy}}{(j-1 \, j \, j+1 \, k) (j\, k-1 \, k \, k+1)},
\ee
where
\begin{align}
\hat{a}_{j1} &= (j\, j+1 \, k \, k+1) (k-1 \, k \, l\, l+1) - (j \, j+1 \, k-1 \, k) (k \, k+1 \, l \, l+1) \,, \notag \\
\hat{a}_{j2} &= -(j-1 \, j \, k \, k+1) (k-1\, k \, l \, l+1) +  (j-1 \, j \, k-1 \, k) (k \, k+1 \, l \, l+1) \,,  \notag \\
\hat{a}_{k1} &= (j-1 \, j \, l \, l+1) (j \, j+1 \, k \, k+1) - (j-1 \, j \, k \, k+1) (j \,  j+1 \, l \, l+1)  \,,  \notag \\
\hat{a}_{k2} &= - (j-1 \, j \, l \, l+1) (j \,  j+1 \,  k-1 \, k) + (j-1 \, j \, k-1 \, k) (j \,  j+1 \, l \, l+1) \,,  \notag \\
\hat{a}_{jk} &= -(j-1 \, j \, j+1 \, l+1) (k-1 \, k \, k+1 \, l) + (j-1 \, j \, j+1 \, l) (k-1 \, k \, k+1 \, l+1) \,,  \notag \\
\hat{a}_{\overline{jk}} &=  -(j \, k \, l \, l+1)\,.
 \end{align}
This can be used to simplify any of the pentaboxes or double pentagons appearing in any two-loop amplitude.

\subsection{Definition of the two-loop integrals}
Here we define the two-loop integrals appearing in (\ref{MHV2loopimproved}).
The double pentagon integral with ``magic numerator'' is given by
\begin{eqnarray}
I_{6;2;m1} &=& \int \frac{ dZ_{AB} dZ_{CD}}{ (i \pi^2)^2} \, \Bigg\lbrack
 \frac{\fbr{13AB} \fbr{46CD} \fbr{6134} \fbr{1245} \fbr{2356} }
 { \mfbr{61AB} \mfbr{12AB} \mfbr{23AB} \mfbr{34AB}}  \times \nonumber \\
 && \hspace{2cm} \times  \frac{1}{\fbr{ABCD} \mfbr{34CD} \mfbr{45CD} \mfbr{56CD} \mfbr{61CD}}  \Bigg\rbrack \,,
 \end{eqnarray}
 and the remaining integrals are defined as
 \begin{eqnarray}
 I_{6;2;m2}  \hspace{-2mm}&=& \hspace{-2mm} \int 
 \frac{dZ_{AB} dZ_{CD} \, (i \pi^2)^{-2} \,  \fbr{36CD} \fbr{6123} \fbr{1234} \fbr{1256} }
 { \mfbr{61AB} \mfbr{12AB} \mfbr{23AB} \fbr{ABCD} \mfbr{23CD} \mfbr{34CD} \mfbr{56CD} \mfbr{61CD}}  \,,
\\
  I_{6;2;m3} \hspace{-2mm}&=& \hspace{-2mm}  \int 
 \frac{dZ_{AB} dZ_{CD} \, (i \pi^2)^{-2} \,  \fbr{35CD} \fbr{6123} \fbr{1234} \fbr{1245} }
 { \mfbr{61AB} \mfbr{12AB} \mfbr{23AB} \fbr{ABCD} \mfbr{23CD} \mfbr{34CD} \mfbr{45CD} \mfbr{56CD} }   \,,
 \end{eqnarray}
 and
 \begin{eqnarray}
  I_{6;2;7} \hspace{-2mm}&=& \hspace{-2mm}  \int 
 \frac{ dZ_{AB} dZ_{CD} \, (i \pi^2)^{-2} \, \fbr{1245} \fbr{6123} \fbr{3456} }
 { \mfbr{61AB} \mfbr{12AB} \mfbr{23AB} \fbr{ABCD} \mfbr{34CD} \mfbr{45CD} \mfbr{56CD}}  \,,
 \\
  I_{6;2;15} \hspace{-2mm}&=& \hspace{-2mm}  \int 
 \frac{ dZ_{AB} dZ_{CD} \, (i \pi^2)^{-2} \, \fbr{1234} \fbr{6123} \fbr{2345} }
 { \mfbr{61AB} \mfbr{12AB} \mfbr{23AB} \fbr{ABCD} \mfbr{23CD} \mfbr{34CD} \mfbr{45CD}}  \,.
 \end{eqnarray}

\subsection{Results for generic kinematics}

In the small $m^2$ expansion we obtain the following results.
\begin{eqnarray}
I_{6;2;m1}  &=& \cO(1) \,,
\\
I_{6;2;m2}  &=&
 -\frac{1}{24} \log^4 m^2 +  \log^3 m^2  \log\left( \frac{x_{13}^2 x_{24}^2  x_{26}^2}{x_{14}^2 x_{36}^2} \right) + \cO( \log^2 m^2)\,,
\\
I_{6;2;m3} &=& \frac{2}{3} \log^3 m^2    \log\left(\frac{x_{25}^2 x_{36}^2}{x_{35}^2 x_{26}^2}\right)  +  \cO( \log^2 m^2 )\,,
\\
I_{6;2;7}  &=& \frac{1}{4} \log^4 m^2 \nonumber \\
&& + \log^3 m^2  \left[ \frac{1}{2} \log\left( \frac{ x_{15}^2 x_{24}^2 x_{26}^2 x_{35}^2}{x_{13}^2  x_{46}^2} \right) - 2 \log(x_{25}^2) \right] + \cO( \log^2 m^2 )\,,
\\
I_{6;2;15} &=&  \frac{7}{12} \log^4  m^2  \nonumber \\
&& +\log^3 m^2 \left[  \frac{2}{3} \log\left( \frac{ x_{25}^2 x_{36}^2 }{x_{13}^2 x_{15}^2 } \right) -\frac{7}{3} \log(x_{26}^2) \right]  +  \cO( \log^2 m^2 )\,.
\end{eqnarray}
Here we used integrals like
\begin{eqnarray}
 \int_{{\rm Re}(z)=-1/2} dz \,x^z \, \Gamma^2 (-z) \Gamma^2 (1+z) = \frac{-1}{1-x} \log(x)\,.
\end{eqnarray}
We also computed the  $\log^2(m^2)$ analytically.
One can check that the infrared divergent terms in (\ref{M62divpart}) are correctly reproduced.

\subsection{Analytic results in the limit $u\to 0$}
Here we give the analytical results for all integrals for $x_{i,i+2}^2 =1\,, x_{i,i+3}^2= u^{-1/2}$
as $u \to 0$ (after having taken the regulator limit $m^2 \to 0$). We denote $U=\log u$ and $L = \log m^2$. 
We obtain
\begin{eqnarray}
I_{6m1} &=&  \frac{1}{4} U^4 + U^2  \frac{\pi^2}{3} +  2 U  \zeta_{3} + \frac{7}{120}\pi^4  + \cO(m^2, u)\,, \\
I_{6m2} &=& -\frac{1}{24} L^4 + \frac{1}{6} L^3 U  + \frac{1}{4} L^2 \left(-U^2 - \pi^2 \right) +\frac{1}{6} L \left( U^3 + 3 U \pi^2 - 12 \zeta_{3}  \right)  \nonumber \\
&& \left( - \frac{1}{24} U^4  - U^2  \frac{\pi^2}{4} -2 U \zeta_{3} - \frac{7}{40} \pi^4 \right)  + \cO(m^2, u)\,, 
\\
I_{6m3} &=& - \frac{2}{3} L^3 U   + L^2 \left(-U^2 - \frac{\pi^2}{3} \right) + 
 L \left(- \frac{1}{2} U^3 -  \frac{\pi^2}{3} U - 4 \zeta_{3} \right) \,, 
 \nonumber \\
&& + \left(  \frac{5}{48} U^4 + \frac{13 \pi^2}{12} U^2 +4 \zeta_{3} U + \frac{17 \pi^4}{45}\right) + \cO(m^2, u)\,,   \end{eqnarray}
 and
 \begin{eqnarray}
I_{7} &=& \frac{1}{4} L^4 + L^3 U - \frac{2 \pi^2}{3} L^2 + L \left(  -\frac{1}{2} U^3 - \frac{2 \pi^2}{3} U + 4 \zeta_{3}  \right)  \nonumber \\
&& +
 \left( \frac{3}{16} U^4 + \frac{2 \pi^2}{3} U^2 -2 U \zeta_{3} + \frac{7 \pi^4}{72}  \right)
 +  \cO(m^2, u)\,, 
\\
I_{15} &=& \frac{7}{12} L^4 - \frac{2}{3} L^3 U - \frac{\pi^2}{6} L^2 + \frac{1}{6} L \left( U^3 + 4 L \pi^2 - 48 \zeta_{3} \right) \nonumber \\
&&  \left( -\frac{1}{48} U^4 - \frac{\pi^2}{6} U^2 + 4 U \zeta_{3} - \frac{\pi^4}{20} \right) +  \cO(m^2, u)\,.
\end{eqnarray}
 Putting everything together, we obtain
\begin{equation}
M_{6}^{(2)}= \frac{9}{8} L^4 + L^2 \left( \frac{9}{8} U^2 - \frac{\pi^2}{8} \right) - 3 \zeta_{3} L   +
\left( \frac{9}{32} U^4 + \frac{\pi^2}{16} U^2 + \frac{\pi^4}{48} \right)
 + \cO(m^2, u) \,,
\end{equation}
and hence
\begin{equation}
\lim_{u\to 0} \cR_{6}^{(2)}(u,u,u) = \frac{\pi^2}{8} \log^2 u +\frac{17 \pi^4}{1440} + \cO(u) \,,
\end{equation}
which is quoted in the main text.

We have similar results for the limit $u \to \infty$.

\section{Box integrals for one-loop MHV amplitudes}
In order to write down the one-loop $n$-point MHV amplitude up to $\cO(m^2)$, we require the so-called ``one-mass'' and
``two-mass easy'' box integrals (in our case with uniform internal mass $m$). They are readily evaluated and are given by
\begin{eqnarray}\label{app-F1m}
F^{1m}(x_{46}^2,x_{15}^2,x_{14}^2) &=& \log^2 \frac{m^2}{x_{46}^2} +  \log^2 \frac{m^2}{x_{15}^2}- \log^2 \frac{m^2} {x_{14}^2} - \log^2 \frac{x_{15}^2}{x_{46}^2} - \frac{\pi^2}{3} \nonumber \\
&& - 2 \, {\rm Li}_{2}\left(1- \frac{x_{14}^2}{x_{15}^2}\right) - 2 \,  {\rm Li}_{2}\left(1- \frac{x_{14}^2}{x_{46}^2}\right) + \cO(m^2)\,.
\end{eqnarray}
and
\begin{eqnarray}\label{app-F2me}
F^{2me}(x_{14}^2,x_{36}^2,x_{13}^2,x_{46}^2)  &=& - \log^2 \frac{m^2}{x_{14}^2} 
 - \log^2 \frac{m^2}{x_{36}^2} + \log^2 \frac{m^2}{x_{13}^2} + \log^2 \frac{m^2}{x_{46}^2} + \log^2 \frac{x_{14}^2}{x_{36}^2} \nonumber \\
&& +2 \,{\rm Li}_{2}\left( 1- \frac{x_{13}^2}{x_{14}^2}\right) +2\, {\rm Li}_{2}\left( 1- \frac{x_{13}^2}{x_{36}^2}\right)+ 
2 \, {\rm Li}_{2}\left( 1- \frac{x_{46}^2}{x_{14}^2}\right)  \nonumber \\
&&+2 \,{\rm Li}_{2}\left( 1- \frac{x_{46}^2}{x_{36}^2}\right)
- 2\, {\rm Li}_{2}\left( 1- \frac{x_{13}^2 x_{46}^2}{x_{14}^2 x_{36}^2}\right) +\cO(m^2) \,.
\end{eqnarray}
Here $F^{1m}= x_{46}^2 x_{15}^2 I^{1m} $ and $F^{2me}= (x_{13}^2 x_{46}^2 - x_{14}^2 x_{36}^2 )  I^{2me}$.
Note that both functions can be identified with their dimensional regularisation counterparts 
(up to a constant in the finite part of $F^{1m}$) when expanding the latter in $\epsilon$ and setting 
the dimensional regularisation scale $\mu = m^2$.

\bibliographystyle{nb.bst}
\bibliography{loopsdraft}

\end{document}